\documentclass[aps,pra,twocolumn,groupedaddress,showpacs,10pt]{revtex4-1}
\usepackage{graphicx}
\usepackage{amsfonts}

\begin{document}


\title{Bichromatic Slowing of Metastable Helium}

\author{M. A. Chieda}
\author{E. E. Eyler}
\affiliation{Physics Department, University of Connecticut, Storrs, CT 06269}

\date{\today}

\begin{abstract}

We examine two approaches for significantly extending the velocity range of the optical bichromatic force (BCF), to make it useful for laser deceleration of atomic and molecular beams.  First, we present experimental results and calculations for BCF deceleration of metastable helium using very large BCF detunings, and discuss the limitations of this approach.  We consider in detail the constraints, both inherent and practical, that set the usable upper limit of the BCF.  We then show that a more promising approach is to utilize a BCF profile with a relatively small velocity range in conjunction with chirped Doppler shifts, to keep the force resonant with the atoms as they are slowed. In an initial experimental test of this chirped BCF method, helium atoms are slowed by $\sim 370$~m/s using a BCF profile with a velocity width of $\lesssim 125$~m/s. Straightforward scaling of the present results indicates that a decelerator for He* capable of loading a magneto-optical trap (MOT) can yield a brightness comparable to a much larger Zeeman slower.
\end{abstract}


\maketitle

\section{Introduction}

Most laser slowing methods rely on the radiative force in a monochromatic field \cite{Metcalf99, Metcalf03, Baldwin05, Barry12}. The momentum transfer $\hbar k$ arising from photon absorption allows atoms to be accelerated or decelerated, and their velocity distribution to be narrowed (cooled). However, the radiative force is limited in its magnitude by the spontaneous emission rate $\gamma$, to a maximum of $F_{\text{rad}} = \hbar k \gamma/2$. Long slowing times allow the small transverse velocity spread in an atomic beam to significantly reduce the source brightness, negatively impacting the loading of the slowed atoms into a magneto-optical trap (MOT).

Unlike the radiative force, the dipole force arising from intensity gradients in a monochromatic standing wave is not limited by the radiative decay rate. However, it alternates sign on the scale of an optical standing wave. The average force is zero, limiting its usefulness in a decelerator, although a pulsed slower can be created by utilizing a transient standing-wave lattice produced by an intense pulsed laser \cite{Barker10}. Another way around this limitation, demonstrated by Kazantsev and Krasnov~\cite{Kazantsev87}, is to rectify the dipole force. Adding a second frequency to the standing wave results in a modulated light shift, which can be adjusted to keep the sign of the dipole force positive. This technique eventually resulted in the deflection of sodium atoms by a few m/s by Grimm and coworkers~\cite{Grimm90}.

Continued refinement of slowing techniques in bichromatic fields by Grimm and coworkers eventually led to the observation of a much larger rectified dipole force in cesium, with a velocity range of 225~m/s~\cite{Soding97}. This was the birth of the optical bichromatic force (BCF), which relies on a pair of counterpropagating two-color beams that, when the intensity and standing wavelength are carefully selected, will coherently drive the atom through cycles of photon absorption and stimulated emission much more rapidly than the radiative decay rate.

These developments led to more extensive research on the bichromatic force, which to date  has been demonstrated in Cs~\cite{Soding97}, Na~\cite{Voitsekhovich89}, Na$_2$~\cite{Voitsekhovich94}, Rb~\cite{Williams99, Williams00}, and metastable helium (He*)~\cite{Cashen01,Cashen02,Cashen03,Partlow04,Partlow04b}.  Despite this progress and the obvious advantages of a strong continuous optical force, there appears to be only one instance in which the BCF has been used with a MOT, a recent experiment in which improved efficiency was demonstrated for atomic beam loading of an $^{87}$Rb MOT \cite{Liebisch12}.

We have designed two experiments to test the potential limits of the BCF and evaluate the feasibility of an atomic beam slower for He*. The first set of experiments tests the sensitivity and limitations of the BCF at large detunings. This is required for a static atomic slower design, which would utilize one or two pairs of bichromatic beams configured such that the BCF profile has a wide velocity range, capable of decelerating He* atoms by several hundred m/s.

The second set of experiments explores an alternative BCF slower scheme that avoids many of the difficulties of using large bichromatic detunings. Here BCF beams with a relatively small detunings are used, with the addition of dynamic frequency chirping to maintain resonance with the changing Doppler shifts of the decelerating atoms. This design conserves laser power at the cost of added complexity in the control electronics.  More importantly, it keeps the BCF parameters in a range within which they are robust against small misadjustments.  We demonstrate deceleration by 370 m/s in a prototype design, constrained mainly by the low-power diode lasers used for initial tests.  We also discuss straightforward improvements that can more than double this range.

\section{Modeling the bichromatic force\label{sec:BCF}}

\subsection{$\pi$-pulse model\label{sec:pipulse}}

\begin{figure}
\includegraphics[width=\columnwidth]{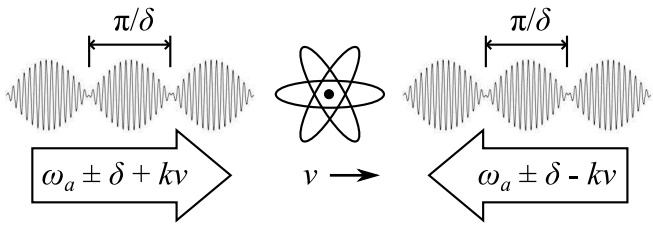}
\caption{Simple $\pi$-pulse model of the optical bichromatic force on an atom, adapted from Fig. 1 of Ref. \cite{Chieda11}. Two-color beams impinge from each direction, offset by frequencies $\pm \delta$ from resonance at $\omega_a$. For large atomic velocities $v$, Doppler offsets of $\pm kv$ must also be added as shown.  The frequency pairs at $\pm \delta$ give rise to a series of rf beat notes, each with an area of approximately $\pi$.}
\label{fig:BCF_concept}
\end{figure}

An intuitive but somewhat inaccurate model for the BCF \cite{Soding97,Voitsekhovich89} is based on the concept of periodic momentum exchange between the atom and the co- and counterpropagating bichromatic beams using $\pi$-pulses. As shown in Fig. 1, each pair of two-color beams is offset by the atomic Doppler shift $kv$ so that in the frame of the moving atom, the frequencies are offset from the atomic resonance $\omega_0$ by  $\pm \delta$.  Thus the atom experiences from each direction a traveling sequence of beat notes, each with duration $T = \pi/\delta$.  In the $\pi$-pulse model, the two sets of beat notes are treated as if they are independent non-overlapping pulse trains.  In this approximation, if the laser power is chosen so that the beat note area satisfies the $\pi$-pulse condition for an atomic transition, the atom will experience a sequence of alternating excitations and stimulated de-excitations.  This occurs when
\begin{equation}
\Omega = \pi\delta/4,
\label{eq:Rabi_pi}
\end{equation}
where $\Omega = \gamma(I/2I_{\text{sat}})^{1/2}$ is the on-resonance Rabi frequency of each bichromatic frequency component \cite{Metcalf03}.  A large decelerating force to the left occurs if the excitations come from the right-hand beam and the de-excitations from the left. If $\delta/\pi \gg \gamma$, this force will greatly exceed the radiative force $F_{\text{rad}}$.

This force is quite tolerant of deviations from the exact $\pi$-pulse condition so long as the left-hand beat notes closely match the right-hand ones.  In the pulse-pair model, the two pulses will then appear to the atom as time-reversed images of one another.  The second pulse in each pair reverses the curve on the Bloch sphere traced out by the first \cite{Milonni10}, leaving the atom back in the ground state after each two-pulse cycle.  This simple time-domain picture is very useful for estimating the breakdown of the BCF at large detunings or large atomic velocities, a subject that we discuss further in Section \ref{sec:dephasing}.

The $\pi$-pulse model also leads to a simple and intuitive estimate for the velocity range of the BCF, which appears to have been overlooked in the previous literature.  To a first approximation, finite atomic velocities merely induce Doppler shifts $+kv$ and $-kv$ for the two pulses in each pair, and since the pulses still appear as time-reversed images there is little effect.  However, the generalized Rabi frequency is also modified \cite{Milonni10}, because the Doppler shifts detune the center or ``carrier" frequency of each beat note from the atomic resonance:
\begin{equation}
\Omega' = \sqrt{\Omega^2 + (kv)^2}.
\label{eq:eff_Rabi}
\end{equation}
If the Doppler shift $kv$ is comparable to the resonance Rabi frequency $\Omega$, the Rabi cycling even for a single pulse is so badly out of phase that the BCF will be disrupted.  In the simplified limit of a rectangular pulse with $\Omega \approx \delta$, the resulting full width of the velocity range is
\begin{equation}
\Delta v \approx 2\frac{\delta}{k},
\label{eq:pipulse_v_range}
\end{equation}
in agreement with much more sophisticated models apart from the factor of 2.

The relative phase between the co- and counterpropagating beat notes is critical to the optimization of the BCF, as can easily be seen by considering a single pair of $\pi$-pulses separated by time $\Delta t$, the first arriving from the right and the second from the left. If the atom is excited by the first pulse but undergoes spontaneous decay before the second pulse arrives, it will start cycling in the wrong sequence, leading to acceleration.  This can be minimized by making $\Delta t$ as short as possible while maintaining the discreteness of the pulses. Further, if the repetition period $T \gg \Delta t$, an atom that is already cycling in the wrong sequence will be more likely to radiate to the ground state after the two-pulse sequence than in between the pulses, placing it back into the correct sequence for deceleration by future pulse pairs.  For short rectangular pulses the optimal configuration would be with minimal delay $\Delta t$, but for beat note trains, the best one can do is to set the relative phase to $\phi=\pi/2$, where $\phi=0$ corresponds to the limiting case of simultaneous arrival, where the pulses fully overlap and there is no net force, and $\phi=\pi$ is the equally ineffective symmetric case where $\Delta t=T/2$. Averaging over several radiative lifetimes with $\phi=\pi/2$, the correct (decelerating) sequence occurs during 3/4 of the interaction time, while the atom is accelerated during the remaining time. Numerical calculations and experiments support this argument, which reduces the average force by a factor of two compared with perfect in-phase cycling. Because the individual $\pi$-pulses exchange momentum increments of $\hbar k$ at rate $\delta/\pi$, the average bichromatic force is
\begin{equation}
F_{\text{b}}=\hbar k \delta/\pi
\label{eq:F_b}.
\end{equation}
This average force exceeds the radiative force by a factor of $2 \delta / (\pi \gamma)$.

\subsection{Doubly-dressed atom model}

A more complete model that still provides some intuitive insight involves dressing the atomic levels with the red- and blue detuned photon field number states. Originally developed by Grimm and coworkers \cite{Grimm94, Grimm96}, this ``doubly dressed'' model results in a ladder of states separated by $\hbar\delta$ and an infinite tridiagonal Hamiltonian matrix that is well described by Metcalf and Yatsenko~\cite{Yatsenko04}. Numerically solving for the eigenstates of the Hamiltonian provides an estimate of the BCF magnitude that agrees with the $\pi$-pulse model, and is also in agreement that the optimal phase is $\pi/4$.  Unlike the $\pi$-pulse model it predicts that the optimal Rabi frequency is given by
\begin{equation}
\Omega_{\text{opt}} = \sqrt{3/2} \, \delta,
\label{eq:Rabi_DDA}
\end{equation}
This result, larger than Eq. \ref{eq:Rabi_pi} by about 56\%, is in excellent agreement both with experiment and more complete numerical calculations.  Applying Floquet theory and using a second-order perturbation treatment, this model also yields an estimate of the velocity range of the force, but this prediction is too small by about a factor of two, probably because the effects of Doppler shifts are not fully incorporated~\cite{Cashen02}.

\subsection{Numerical solutions of the optical Bloch equations\label{sec:OBEs}}

\begin{figure}
\includegraphics[width=0.8\columnwidth]{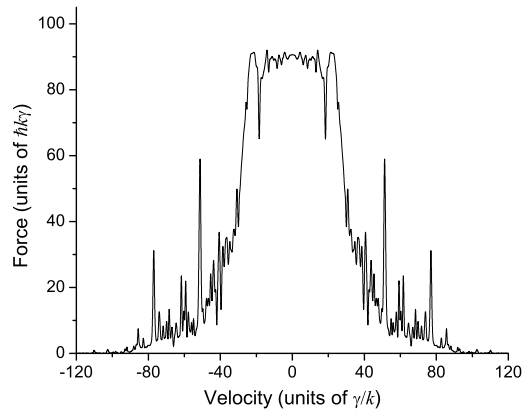}
\caption{Calculated bichromatic force vs. velocity for $\delta = 154\gamma$ and phase $\phi=\pi/2$, for $\Omega_{\text{opt}}= \sqrt{3/2} \, \delta$. This optimum Rabi frequency yields the largest central force, in agreement with Eq. \ref{eq:Rabi_DDA} rather than Eq. \ref{eq:Rabi_pi}. Note that the force magnitude is shown as a multiple of $\hbar k \gamma$, which is twice the radiative force $F_{\text{rad}}$.}
\label{fig:BCF_Profile}
\end{figure}

A more comprehensive but less transparent approach is to numerically solve the optical Bloch equations (OBEs)~\cite{Feynman57,Allen87} in the bichromatic field. The OBEs fully describe the internal state of a two-level atom in the rotating-wave approximation (RWA), taking into account the Rabi cycling of atomic states in the field as well as spontaneous emission, which is added phenomenologically. We obtain numerical solutions using a program based largely on code developed for earlier BCF studies by the authors of Refs.~\cite{Grimm94,Grimm96,Soding97}. The rotating frame is chosen to be at the center frequency of the bichromatic spectrum, leaving some residual time dependence due to the approximately symmetric detunings $\omega_0 \pm \delta$.  The OBEs are solved over small time steps in the bichromatic field and the instantaneous force is evaluated using Ehrenfest's theorem~\cite{Metcalf03},
\begin{equation}
F(z,t)=\hbar[u(t)\nabla\Re(\Omega_r(z,t))-v(t)\nabla\Im(\Omega_r(z,t))],
\end{equation}
and then averaged over several radiative lifetimes. The atomic velocity is not explicit in the force calculation, but is inferred using a constant-velocity approximation so that the atomic position is given by $z = vt$. This process is repeated over a range of atomic velocities to produce force profiles like the one shown in Fig.~\ref{fig:BCF_Profile}.

At the center of the force profile, the calculated amplitude is maximized when the Rabi frequency satisfies Eq. \ref{eq:Rabi_DDA} and the bichromatic beat phase offset is $\phi=\pi/2$, verifying both of the central predictions of the doubly-dressed atom model.  Both predictions also agree well with experimental results.  Under these optimal conditions the maximum force is $F_{\text{BCF}}\approx \hbar k \delta/\pi$, also consistent with both the simplified models and experiments.

The velocity range of the force in the profile of Fig. \ref{fig:BCF_Profile} is about $\delta/2k$, but the profiles take on distinctly different shapes for $\delta \leq \Omega < \sqrt{3/2}\,\delta$ \cite{Chieda12}, developing broad peaks near $\pm\delta/2k$ with a force nearly as large as for the central zero-velocity peak. Atoms traversing a Gaussian bichromatic beam profile will interact with a range of intensities, so a realistic simulation  should include an average over the force profiles.  This yields an improved estimate of the BCF velocity range~\cite{Partlow04,Partlow04b},
\begin{equation}
\Delta v \approx \delta/k,
\end{equation}
that is half as large as the rough estimate in Eq. \ref{eq:pipulse_v_range}.

The time required to slow an atom across this full velocity range is $t_{\text{BCF}} = m \Delta v/F_{\text{BCF}}$. This ``bichromatic slowing time" is independent of the detuning $\delta$ and for helium its value is $t_{\text{BCF}} = 5.9~\mu$s. It follows that even a fast beam of metastable helium atoms, with $\bar{v} \sim 1000~\text{m/s}$, can be decelerated by $\Delta v$ in an interaction length of less than 1~cm.

Another feature of the BCF profile is the sharp drop-off at either end of its range, clearly visible in Fig.~\ref{fig:BCF_Profile}. If the force is applied for a time greater than the bichromatic slowing time, atoms within the velocity range of the force will ``pile up'' at the low-velocity limit. This can yield significant cooling in addition to deceleration, as can be seen in Fig.~\ref{fig:LDS_Results} and Refs.~\cite{Soding97, Cashen02, Partlow04}. Recently Harold Metcalf has argued that the cooling is due to radiation exchange between the bichromatic fields, and may not be subject to limitations tied to the rate of spontaneous emission~\cite{Metcalf08}.

\section{Large-detuning BCF\label{sec:LDS}}

A practical atomic beam slower for metastable helium must reduce the atomic velocity by at least 700--800 m/s to produce a large flux of atoms at MOT capture velocities.  Ideally, this could be achieved with a single pair of counterpropagating beams producing a BCF profile with a correspondingly large velocity range. However, this would require unreasonably high irradiances, so it is necessary to break up the required velocity range in some way.  In this section we investigate the feasibility of a static two-stage slower and more generally, we study the large-detuning limits of the BCF.
\subsection{Experiment\label{sec:LDS_Exp}}

\begin{figure*}
\includegraphics[width=0.8\textwidth]{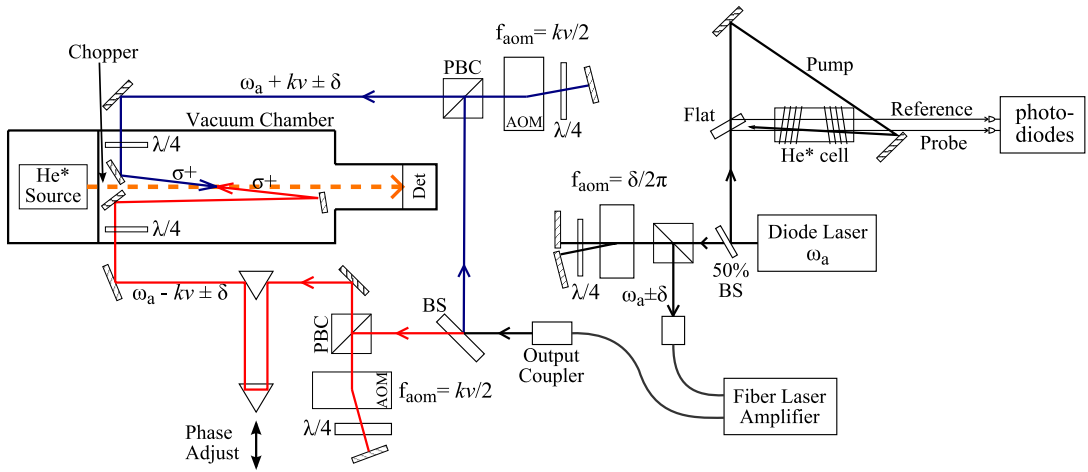}
\caption{(color online).  Experimental configuration for observing the BCF in atomic He using detunings of $123\gamma-278\gamma$ with a static Doppler shift centered at atomic velocity $v$. Polarizing beamsplitter cubes are denoted by PBC, and a beamsplitter by BS.}
\label{fig:LDSexplayout}
\end{figure*}

For efficient deceleration by two successive BCF interactions, the first stage should have a velocity profile Doppler shifted so that its leading edge is located in an appropriate portion of the slow tail of the atom velocity profile, and the second stage should have a smaller Doppler shift with its leading edge just overlapping the trailing edge of the first profile. We have constructed a decelerator in the first-stage configuration to test the concept, as diagrammed in Fig.~\ref{fig:LDSexplayout}. The metastable helium source is a reverse-pumped, liquid nitrogen cooled dc discharge source patterned closely on a prior design \cite{Kawanaka93,Mastwijk98}, with external modifications to permit installation in an existing vacuum chamber.  The metastable atom flux from the source is about $3\times10^{13}$~He* atoms/sr$\cdot$s with a most probable velocity of approximately 1050~m/s and a metastable fraction of roughly $5\times10^{-5}$~\cite{Metcalf99}. The atomic beam passes from the source chamber through a $500~\mu$m diameter skimmer aperture into a time-of-flight chamber.  After passing through a $70~\mu$m diameter collimating aperture, it is mechanically chopped by a tuning fork chopper operating at 160~Hz with a $100~\mu$m slit width.  The resulting chopped atom beam, with a divergence half-angle of 4.1~mrad, travels through the counterpropagating bichromatic decelerating laser beams for a few centimeters. After an additional time-of-flight (TOF) path, the chopped beam impinges on a stainless steel Faraday cup. Electrons ejected by He* atom impacts are collected and detected by a Ceratron continuous dynode electron multiplier. The He* velocity distribution is then calculated from the TOF spectra, with a velocity resolution of about 67~m/s.

A limitation of this method is that it cannot measure velocities slower than about 350 m/s because the slowed atoms will then be overlapped by atoms from the succeeding chopper pulse.  For the experiments reported here, we avoid this problem by selecting an initial velocity range high enough in the velocity distribution that the slowed atoms remain above this lower limit.  To measure the velocities in a practical MOT loading scheme, a modified beam chopper or a laser tagging method would be required.

The bichromatic laser beam pairs are generated from a single diode laser (Toptica DL100) locked to the $2\,^{3}S_{1} - 2\,^{3}P_{2}$ transition at 1083.3~nm using a saturated absorption spectrometer. This laser at the atomic frequency $\omega_a$ is double-passed through an acousto-optic modulator (AOM) operating at the bichromatic detuning frequency $\delta/2\pi$, producing two superimposed frequencies at $\omega_a \pm \delta$.  It is then coupled into a cw fiber laser amplifier (Nufern NuAmp) with a single-mode output of up to 7~W. The amplified beat-note train is split into equal parts, then Doppler frequency shifts of $\pm kv/2 \pi = \pm800$~MHz are added using additional AOMs, to center the bichromatic force profile at a velocity of about 866~m/s. An optical delay line is used to set the relative phase between the two bichromatic beat note trains to the optimal $\phi = \pi/2$~\cite{Soding97}. The bichromatic beams enter the chamber linearly polarized, pass through quarter-wave retarders to produce $\sigma^{+}$ light, and cross the helium atomic beam at an angle of about $1^{\circ}$. The beams are tightly focused to waists with top-hat radii of 0.32~mm to provide the required irradiance without exceeding the capabilities of the amplifier.  Because of this tight focus the overlap with the atomic beam is imperfect, resulting in an interaction region about 3.7~cm long but covering only about 40\% of the atomic beam.

\subsection{Results and analysis\label{sec:LDS_Res}}

\begin{figure}
\includegraphics[width=\columnwidth]{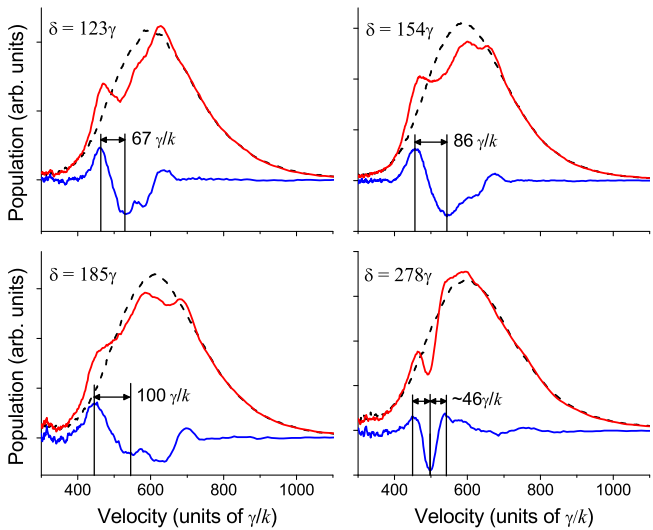}
\caption{(color online). Experimentally observed velocity profiles for BCF slowing, increasing in detuning from top to bottom. The dotted trace (black) is the unslowed distribution, the upper solid trace (red) is the slowed distribution, and the lower solid trace (blue) is the difference. Notice the anomalous result at $\delta = 278\gamma$, consistent with previous observations of an upper limit to the BCF detuning.}
\label{fig:LDS_Results}
\end{figure}

The results for this large-detuning slowing are shown in Fig.~\ref{fig:LDS_Results}.  We extend previous work using a fixed detuning of $185\gamma$ ($\sim$300 MHz) \cite{Cashen03, Partlow04} by exploring several different detunings.  The figure shows raw experimental data without enhancement to compensate for imperfect beam overlap, as was done in some prior BCF experiments on He*~\cite{Partlow04, Partlow04b}. Consequently, the depth of the holes in the velocity profiles can be at most 40\% of the population, which may be further reduced because the atomic beam contains some atoms in the $2\,^1S_0$ metastable state that cannot be slowed.

Unfortunately these experiments were terminated abruptly by the failure of our 7~W fiber amplifier.  As a result, only one set of data was acquired in the previously unexplored range of detunings between $185\gamma$ (300 MHz), where a large BCF was observed \cite{Cashen03}, and $368\gamma$, where no effect could be found \cite{Partlow04}. Our new results are consistent with these prior reports, showing a badly disrupted BCF at an intermediate  $278\gamma$, establishing an appareent upper limit to the BCF that we discuss further in Sec.~\ref{sec:dephasing}.

The changes in the velocity profiles shown in Fig.~\ref{fig:LDS_Results} were confirmed to be due to the BCF by adjusting the beat note phase, by blocking one of the two bichromatic beams, and by blocking a single frequency component in one of the beams. As expected, when the phase is increased to $\phi=3\pi/2$, the sign of the force changes and the deceleration becomes an acceleration. This is perhaps the most obvious and identifying signature of the BCF. When the phase is set to $\phi=0$ or $\phi=\pi$, the force vanishes altogether, also as expected. If one or more of the frequency components is blocked the BCF vanishes and smaller peaks appear in the atomic velocity distribution due to the radiative force from each bichromatic frequency component.

To establish a consistent convention for measuring the extent of slowing and the velocity range of the BCF, we define ``slowing" as the velocity difference between the peak of the slowed atom distribution and the corresponding minimum of the hole in the slowed velocity distribution, as indicated in Fig.~\ref{fig:LDS_Results}. The velocity range of the BCF can be estimated by measuring the full width between the half-maximum of the slow side of the slowed atom peak and the half-minimum on the fast side of the hole in the original velocity distribution. This is the convention used by the Metcalf group in their BCF slowing experiments~\cite{Partlow04, Partlow04b}. In some circumstances this may differ from the actual full width of the BCF velocity profile, but we will adopt this convention to facilitate comparisons between prior results and ours.

\begin{table}%
\caption{\label{tbl:LDSresults}Summary of BCF slowing results, using the measurement conventions described in the text. Note that the result for $\delta = 278\gamma$ does not exhibit a velocity profile consistent with unidirectional BCF slowing, so this slowing measurement is shown only for comparison.}
\begin{ruledtabular}
	\centering
		\begin{tabular}{c c c c c}
			Detuning & \multicolumn{2}{c}{Slowing} & Velocity Range\\
			$\gamma$ &  $\gamma/k$ & $\delta/k$ & $\delta/k$\\
			\hline
			123 & 67 & 0.55 & 1.04\\
			154 & 86 & 0.56 & 1.08\\
			185 & 100 & 0.54 & 1.00\\
			278 & 46 & 0.16 & N/A\\
		\end{tabular}
\end{ruledtabular}	
\end{table}

Examining the slowing data summarized in Table~\ref{tbl:LDSresults}, we verify that the BCF slowing is proportional to $\delta$, and that the velocity range of the force is very close to $\delta/k$ as predicted by the numerical calculations described in Section~\ref{sec:BCF}.  This agrees with previous results from the Metcalf group at $\delta=185\gamma$~\cite{Partlow04}.

Because the length of the slower is about 3.7~cm, the atoms experience 6--8 BCF slowing time intervals $t_{\text{BCF}}$, depending on the detuning. This complicates a direct measurement of the BCF magnitude since the atoms will be slowed by the full velocity range even when the force is well below the optimal value of $\hbar k \delta/2$. Attempts to measure transient deceleration by pulsing the bichromatic beams were not successful because of the limited time resolution of the chopped-beam TOF scheme.

There is much insight to be gained from the anomalous velocity profile obtained with the largest BCF detuning of $\delta=278\gamma$ (450 MHz). The symmetric displacements in the velocity distribution, with a central hole and peaks on either side, indicates both deceleration and acceleration away from a center velocity.  The observed velocity range is much less than expected for the normal BCF at this detuning. As mentioned previously, it was not possible to directly measure the acceleration or deceleration, but only their accumulated effects along the length of the slowing region.

The center of the hole is at the center of the force profile, which for this measurement is Doppler shifted by 800~MHz, or 493~$\gamma/k$. It cannot be caused by any of the individual bichromatic frequency components acting alone because each would have a large Doppler shift located at either $\pm 158\:\gamma/k$ from the center of the velocity hole. We also confirmed that the effect is bichromatic in nature by individually blocking one or more of the bichromatic frequency components, which resulted in a complete loss of the velocity shifts.

It thus appears that a bichromatic force is present at $278\gamma$, but that it is no longer a rectified force.  This suggests that the sign of the force is changing sometime during the many stimulated cycles experienced by the atoms in between successive radiative decay events that reset the phase of the cycling.  Indeed, we were able to simulate the measured velocity profile quite well with a numerical Monte Carlo simulation that assumes a randomly directed force with a reduced magnitude that is typical of the BCF at $61\gamma$ \cite{Chieda12}.  This supports a dephasing of the BCF, along with a reduction in magnitude caused either by repeated sign changes or other effects.  The previously observed vanishing of the force at $368\gamma$ suggests that the disruption of the BCF becomes even more severe at still larger detunings. In the next section we explore some possible explanations.

\subsection{Upper limits to the BCF\label{sec:dephasing}}

We have examined a wide variety of physical and engineering limitations that could constrain the largest usable detuning for the BCF \cite{Chieda12}, and here we describe the limiting factors most likely to arise in actual practice.  One obvious issue is that the required laser power scales with $\delta^2$, setting an engineering limitation.  In terms of the saturation irradiance $I_s$, the required irradiance from each beam (Eq. \ref{eq:Rabi_DDA}) is
 \begin{equation}
I_b = 3 I_s \left(\frac{\delta}{\gamma}\right)^2.
\end{equation}
A typical value of $I_s$ is a few mW/cm$^2$, although for He* it is unusually small at 0.17 mW/cm$^2$.  Obtaining the necessary power is thus a significant challenge for cw experiments with detunings of several hundred times the natural linewidth $\gamma$.

Another major consideration is the impact of imperfections that cause cumulative dephasing of the stimulated cycling.  The time-domain $\pi$-pulse model of Section \ref{sec:pipulse} is particularly useful for estimating the size of these effects.  The self-cancellation of each pulse pair depends on precise matching of the left-hand and right-hand beat note ``pulses" in Fig. \ref{fig:BCF_concept}.  If they are not symmetric, each successive pulse pair leaves the atom with a Bloch vector having a larger phase angle, and when the accumulated phase change reaches $\pi$, the direction of the force is reversed.  To make approximate estimates, we can replace the bichromatic beat notes with rectangular pulse trains, each with a repetition period $T = \pi/\delta$ and pulse duration $T/2$, with their arrival times offset by $T/4$.  If the resonant Rabi frequencies are $\Omega_1$ and $\Omega_2$ for pulses arriving from the left and the right, the total accumulated phase error is approximately
\begin{equation}
\frac{1}{2}\int_{0}^{t'} \left(\sqrt{\Omega_1^2 + (\omega_1'(t))^2} - \sqrt{\Omega_2^2 + (\omega_2'(t+T/4))^2}\, \right) dt.
\label{eq:dephasing}
\end{equation}
Here the generalized Rabi frequencies incorporate possible detunings $\omega_1'$ and $\omega_2'$ of the bichromatic center frequencies from the atomic transition frequency, as in Eq. \ref{eq:eff_Rabi}.  The integration time $t'$ is the interval between successive resets of the stimulated cycling by radiative decay or collisional quenching to the ground state.  It is given by $t' = 4/\gamma$ for an ideal BCF cycle in which the atoms spend 1/4 of their time in the excited state.

The requirement for a well-defined BCF is that the dephasing in Eq. \ref{eq:dephasing} is less than $\pi$.  Because the Rabi frequency is proportional to the detuning, this becomes increasingly difficult to achieve at large detunings.  Three possible sources of dephasing are:

1. Intensity imbalance between the left-hand and right-hand bichromatic beams.  This appears as a difference between $\Omega_1$ and $\Omega_2$ and leads to a first-order phase error.  Setting the detunings to zero yields the condition $\Omega_1 - \Omega_2 < 2 \pi/t'$.  For a nominal Rabi frequency of $\Omega = 2 \delta$ that yields a $\pi$ pulse in this approximation, the corresponding fractional imbalance is
\begin{equation}
\frac{\Omega_1 - \Omega_2}{\Omega} \lesssim \frac{\pi}{\delta\, t'}, \mbox{ or } \frac{I_1 - I_2}{I} \lesssim \frac{2\pi}{\delta\, t'},
\label{eq:imbalance}
\end{equation}
using the proportionality of the irradiance $I$ to $\Omega^2$.  In a badly disrupted BCF cycle the probability of atomic excitation will be close to the saturated statistical value of 1/2, so we set $t' \simeq 2/\gamma$ in Eq. \ref{eq:imbalance}.  This yields predictions generally consistent with experimental experience, although prior work on cesium suggests that the sensitivity to beam imbalance is overestimated by about a factor of two \cite{Soding97}.  For He*, Eq. \ref{eq:imbalance} predicts that a detuning of $\delta/2\pi$=250 MHz requires irradiances balanced within 2\%, and at 500 MHz, within 1\%.  In practice beam balancing beyond the 1\% level is exceedingly difficult because of the imperfect spatial profile of the beams, accounting in part for the experimental observation of efficient slowing at $185\gamma$ (300 MHz) but not at $278\gamma$ or above.

Interestingly, numerical OBE simulations predict much less sensitivity to imbalance at large detunings than the $\pi$-pulse model and experiments.  This appears to be because the calculated force profile is dominated under these conditions by multiphoton``Doppleron'' resonances \cite{Bigelow90,Tollett90}, and their effects are almost certainly overestimated because our model assumes perfect two-level atoms that experience a uniform laser irradiance.

2. ``Decelerative dephasing" due to rapidly changing Doppler shifts as the atoms are slowed.  A constant bichromatic force $F_b$ given by Eq. \ref{eq:F_b} will cause a time-dependent velocity change $\Delta v(t) = -F_b\,t/m$, leading to detunings in Eq. \ref{eq:dephasing} of $+k \Delta v(t)$ and $-k \Delta v(t+T/4)$.  In Ref. \cite{Chieda12} a large effect from these detunings is estimated because the subtraction of the two terms in Eq. \ref{eq:dephasing} is not fully taken into account. However, after this subtraction only a very small quadratic term depending on $T/4$ remains, and no significant effect is predicted even for helium with its very small mass.  Appreciable dephasing would occur only if there were also an overall shift $\omega_0$ of the bichromatic center frequency from resonance.  Then the detunings would become $\omega_1' = \omega_0 +kv(t)$ and $\omega_2' = \omega_0 - kv(t+T/4)$, yielding a term proportional to $\delta$.  Asymmetric detunings $\pm kv$ due to Doppler shifts, even if large, do not cause this problem because they retain the mirror-image balance of pulses from the left and from the right.  It is unclear whether this cancellation is equally complete in the actual BCF configuration with its partially overlapping beat-note pulses, but major modifications to our present numerical modeling will be necessary to obtain a definitive answer.

3. rf phase errors. An rf phase shift between the right-hand and left-hand BCF beams will alter the nominal $\pi/2$ phase shift between successive beat notes. Phase errors can arise from external factors such as phase noise in the electronics, but also from the finite spatial extent of the rf beat notes.  At 300 MHz the length of an rf beat note is 50 cm, which is much longer than the 4 cm slowing region, but nevertheless leads to to an rf phase phase deviation of 14.4$^\circ$ between the beginning of the region and its end.  Numerical modeling of the OBEs is necessary to analyze the effects, because the quantitative dependence on the rf phase is not included in the $\pi$-pulse model.  A full analysis is given in Section 3.5.2 of Ref. \cite{Chieda12}, where it is shown that the BCF has a full width at half-maximum of about 9$^\circ$ in phase at 300 MHz detuning, or 5$^\circ$ at 600 MHz.  Thus as $\delta$ increases, the sensitivity to phase increases even as phase variations within the slowing region grow in proportion to $\delta$, and it appears there is little to gain from using a detuning much beyond 300 MHz.  While this problem does not actually reverse the BCF deceleration, it reduces the force enough to make a single-stage slower for metastable helium physically impossible.

A final consideration is that at large irradiances, atomic energy levels can be modified by ac Stark shifts due to laser-induced coupling to distant states.  For helium this is not a significant concern, as is shown in detail in Section 3.2 of Ref. \cite{Chieda12}.  However, in other atomic and molecular systems this could become a critical issue, since it not only leads to symmetric shifts $\omega_1'$ and $\omega_2'$ in Eq. \ref{eq:dephasing}, but also to rapid variations with time and position due to the overlapping traveling BCF beat notes.  A brief discussion in the context of molecular slowing is given in Ref. \cite{Chieda11}.

Although we could not obtain sufficient experimental information at large detunings to definitively establish the cause of the anomalous result at $\delta = 278 \gamma$, it is clear that both intensity imbalance dephasing and rf errors are predicted to cause major problems at this detuning.  To achieve velocity changes much larger than those observed at $\delta = 185 \gamma$, a different approach is needed.

\section{Chirped BCF\label{sec:Chirp}}

The chirped BCF decelerator accomplishes the same goal of slowing metastable helium atoms using bichromatic forces without requiring the large bichromatic detunings $\pm \delta$ of the static slower described in Sec.~\ref{sec:LDS_Exp}. Instead we substitute a BCF with a relatively small detuning and thus a limited velocity range, in which the center frequency of each bichromatic beam pair is dynamically adjusted to stay resonant with the He* atoms while they are slowed. This is done by linearly chirping~\cite{Watts86, Sheehy89} both laser frequencies to follow the changing Doppler shift. This greatly reduces the laser power requirements and allows the slower to operate well within the range of detunings for which the BCF is fully effective and reliable.

\subsection{Modeling\label{sec:ChirpModel}}

\begin{figure}
\includegraphics[width=\columnwidth]{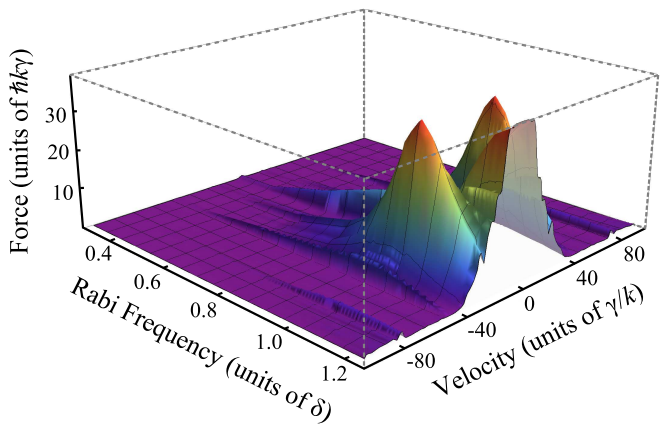}
\caption{(color online). Three-dimensional BCF surface used in the Monte Carlo chirped BCF models. The beat note phase is assumed constant over the interaction region so that the force is a function only of the atomic velocity and Rabi frequency.  The surface is interpolated from numerical solutions to the OBEs. The force is zero at Rabi frequencies less than $0.4\delta$ and absolute velocities greater than $100\,\gamma/k$.}
\label{fig:3D_BCF}
\end{figure}

To support our experiments, we have extended the numerical OBE calculations of Sec. \ref{sec:OBEs} by using the force profiles to create a Monte Carlo model of chirped BCF slowing.  This allows us to predict the performance as as a function of the chirp magnitude and rate.  To accomplish this, a two-parameter force function $F(v,\Omega)$ is needed because both the atomic velocity and the Rabi frequency evolve slowly with time.  The required profile is interpolated from a series of BCF profiles calculated for Rabi frequencies throughout the range from zero to $\sqrt{3/2}\,\delta$.  The resulting BCF surface, depicted in Fig.~\ref{fig:3D_BCF}, shows a pair of prominent peaks at $\Omega \approx 1.0\,\delta$ in addition to the primary maximum at $\Omega = \sqrt{3/2}\,\delta$.

Initial atomic velocities at the start of the Monte Carlo simulation are assigned at random using a probability distribution function derived from an empirical velocity distribution. The atoms are also assigned random initial time offsets between $t=0$ and $-40~\mu$s to account for the finite atomic pulse duration, where $t$=0 corresponds to the beginning of the chopper aperture transit across the atomic beam path. The progress of an individual atom proceeds in small time steps and the change in velocity due to interaction with the bichromatic beams is evaluated at each step.  The Monte Carlo model also accounts for the finite length of the chopped atomic beam pulse, the gaussian bichromatic beam profile, the relative sizes of the atomic beam and bichromatic beam waists, and the transit time of the individual atoms into and through the bichromatic beams.  A fuller account is given in Ref. \cite{Chieda12}.

Once all the atoms have passed through the interaction region, the time of flight for each atom is evaluated and the distributions of both the TOF and the final velocity are calculated to simulate experimental results.  Our initial results based on the full magnitude of the optimal BCF predicted that linear chirp ramps with durations of $10-20~\mu$s would work well, with little change in the size and shape of the slowed atom peak for chirps with magnitudes as large as 600~MHz. However, as we discuss in Sec.~\ref{sec:Chirp_Exp}, in our experiments the optimal chirp durations were found to be in the range of $30-50~\mu$s, with a clear dependence on the chirp magnitude. This was easily accounted for by reducing the effective BCF magnitude by a factor of two to approximately account for experimental imperfections, at which point the Monte Carlo model predictions (shown in Fig.~\ref{fig:MCmodel}) match the experimental results quite well.

\subsection{Experiment\label{sec:Chirp_Exp}}

\begin{figure}
\includegraphics[width=\columnwidth]{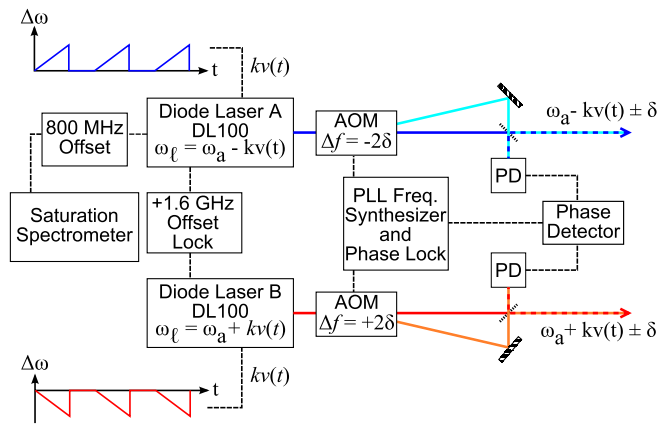}
\caption{(color online). Chirped BCF slower experiment block diagram showing key optical and electronic components. The configuration of the slower beams and atomic beam are the same as shown in Fig.~\ref{fig:LDSexplayout}}
\label{fig:Chirp_Layout}
\end{figure}

The experimental layout for the chirped BCF experiments is shown in Fig.~\ref{fig:Chirp_Layout}, highlighting differences from the large-detuning layout in Fig.~\ref{fig:LDSexplayout}. The chirped slower uses two Toptica DL100 diode lasers, one for the copropagating $+kv$ beam and the other for the counterpropagating $-kv$ beam. This allows independent control of the co- and counterpropagating beam ($\pm kv$) Doppler shifts, a necessity for Doppler chirping. A weak sample beam from the $-kv$ laser is frequency shifted by $+kv$ using an AOM and sent to a saturated absorption spectrometer where it is locked to the $2\,^3S_1 \rightarrow 2\,^3P_2$ transition in helium. The other laser is stabilized at $+kv$ by an offset locking scheme, in which the heterodyne beat note between the two lasers is measured on a photodiode, and locked by a microcontroller based phase locked loop (PLL) circuit to an offset of $+2kv$, typically corresponding to about 1.6 GHz. The lasers are initially locked (without a chirp in progress) at the Doppler shift of a velocity group $v(t_0)$ near the center of the atomic velocity distribution, $\omega_{\ell} = \omega_a \pm kv(t_0)$, and then chirped in opposite directions to follow the atomic Doppler shift as a function of time. Frequency modulation of the two lasers is accomplished using a manufacturer-supplied model DL-MOD interface in one laser and a homemade copy in the other.  The modulation is produced by an FET connected in parallel with the laser diode. A voltage ramp applied to the FET gate causes a portion of the laser diode current to be diverted to ground, changing the laser output frequency.

The bichromatic frequencies are generated for each laser separately using single-passed AOMs driven at an rf frequency $2\delta$. For each laser, the zero- and first-order components are recombined using a 50\% beam splitter to form the bichromatic beam. The beat note phase is controlled by locking the rf phase between a pair of homemade frequency synthesizer boards that drive the two AOMs.   The microcontroller-based synthesizers share a common 10~MHz clock, and a small offset current is added to one of the PLL charge pump outputs to control the phase shift as described in Refs.~\cite{Eyler11,Eyler11b}.

 We found that the rf phase was much less stable than expected because of microphonic motion of the lasers relative to the AOMs, which causes phase variations because of the short acoustic wavelength of the sound waves in the modulators.  A feedback loop was added to provide additional phase stabilization by monitoring the relative rf phase of the optical beat notes using an Analog Devices AD8032 phase detector. By using this phase measurement as an error signal in the PLL phase lock circuit, the RMS phase jitter was reduced to approximately $4^{\circ}$. However, frequent large phase excursions of up to $\pm28^{\circ}$ could not be corrected and reduced the effective magnitude of the BCF.  A better solution might be to utilize a single AOM for both lasers, and this change is planned for future work.

 Our initial experiments were constrained to a very modest BCF detuning of $74\gamma$ because of the limited optical power available from the DL100 diode lasers, which provided about 40 mW in each bichromatic beam pair after accounting for losses in the optics.  For the same reason, the beams were focused to a top-hat radius of 440 $\mu$m that was somewhat smaller than the atomic beam, limiting the fraction of atoms that could be slowed.  This situation could be greatly improved by the addition of a medium-power optical amplifier, such as a tapered laser amplifier diode.

\subsection{Results and analysis\label{sec:Chirp_Res}}

\begin{figure}
\includegraphics[width=\columnwidth]{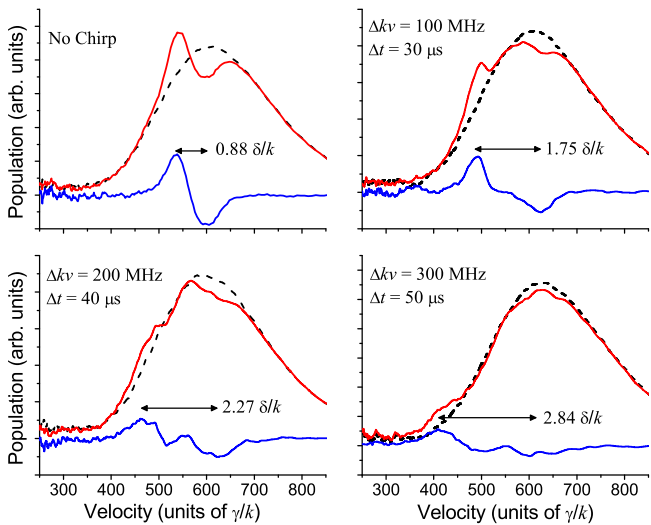}
\caption{(color online). Experimentally observed velocity profiles for frequency-chirped BCF slowing with $\delta=74\gamma$, $\Omega_r=\sqrt{3/2}\,\delta$, and $\phi=\pi/2$.  The four panels show chirp ramps of by 0~MHz, 100~MHz in 30~$\mu$s, 200~MHz in 40~$\mu$s, and 300~MHz in 50~$\mu$s.}
\label{fig:Chirp_Results}
\end{figure}

The measured He* velocity distributions for chirped BCF slowing using a detuning of $74\gamma$ and chirp magnitudes up to 300~MHz are shown in Fig.~\ref{fig:Chirp_Results}. We selected a relatively high initial velocity range centered at 800 m/s only because our present velocity measurement scheme is unsuitable for atoms slowed below 350 m/s, as described in Sec. \ref{sec:LDS_Exp}. The results show the predicted increase in slowing with chirp magnitude, indicating that the chirped BCF profile remains resonant with a fraction of the atoms while they are slowed.

An analysis of these chirp results is shown in Table~\ref{tbl:ChirpResults}, in which the measurements of slowing and velocity range are defined in Sec.~\ref{sec:LDS_Res}. A detuning of only $\delta=74\gamma$ with a 300~MHz frequency chirp provides a measured slowing of 210~$\gamma/k$ or 370 m/s, the largest amount of slowing reported in any BCF experiment to date.  This is more than twice the slowing measured for a static detuning of $\delta=185\gamma$ as reported in Table~\ref{tbl:LDSresults}. The laser irradiance requirements are lower by nearly a factor of ten than what would be required for a static slower with the same velocity range, even if the problems outlined in Sec. \ref{sec:dephasing} could be overcome.

\begin{table}
\caption{\label{tbl:ChirpResults}Summary of experimental BCF chirp results for $\delta=74\gamma$. All experiments used the optimal BCF parameters $\Omega_r=\sqrt{3/2}\,\delta$ and $\phi=\pi/2$.  Slowing is reported both in units of $\gamma /k$ = 1.76 m/s and $\delta /k$ = 130 m/s to facilitate comparison with static BCF deceleration.}
\centering
\begin{ruledtabular}
	\begin{tabular}{c c c c c c}
		Chirp & Ramp & \multicolumn{2}{c}{Slowing} & Velocity Range \\
		MHz & $\mu$s & $\gamma/k$ & $\delta/k$ & $\delta/k$ \\
		\hline
		0 & 0 &65 & 0.88 & 1.57 \\
		100 & $20-40$ & 129 & 1.75 & 2.58 \\
		200 & $40-50$ & 168 & 2.27 & 3.01 \\
		300 & 50 & 210 & 2.84 & 3.67 \\
	\end{tabular}
\end{ruledtabular}
\end{table}

The optimal experimental chirp ramp durations are found to increase from about 30 to 50 $\mu$s as the chirp magnitude increases from 100-300 MHz.  As mentioned previously, the Monte Carlo model described in Sec.~\ref{sec:ChirpModel} yields results in good agreement if the bichromatic force magnitude is reduced by a factor of two from its ideal value. This reduction could easily be caused by experimental imperfections such as rf phase jitter and imperfect gaussian beam profiles.

\begin{figure}
\includegraphics[width=\columnwidth]{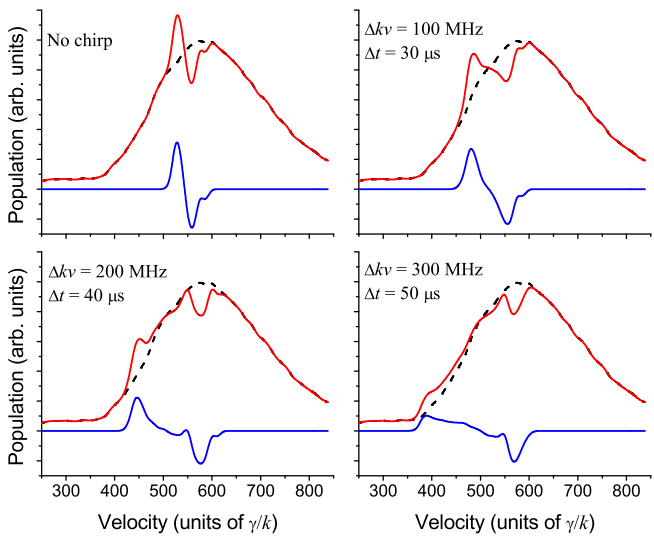}
\caption{(color online). Chirped BCF Monte Carlo model output after reducing BCF magnitude, using the same experimental parameters used to obtain the results in Fig.~\ref{fig:Chirp_Results}}
\label{fig:MCmodel}
\end{figure}

The full velocity profiles predicted by these model calculations are shown in Fig.~\ref{fig:MCmodel}.  Again the agreement with experiment is good.
 The most probable velocity of the slowed group of atoms experimentally matches the model prediction within 20~$\gamma/k$, although the fraction of slowed atoms is relatively small in this initial experimental configuration.  There was no slowing observed when the chirp was increased to 400~MHz, indicating badly sub-optimal BCF conditions. The model predicts this behavior in two scenarios---a smaller BCF magnitude than expected, or periodic, temporary loss of the BCF as would be caused by occasional large phase errors. Both of these factors were potentially present in the experimental configuration for this initial test.

The small fraction of slowed atoms was caused not only by limited beam overlap, but by practical difficulties in using a chopped atomic beam to measure velocity distributions with a single frequency-chirp ramp.  The atomic beam pulse duration, about 40~$\mu$s at the chopper, expands to more than 120~$\mu$s by the time the atoms reach the interaction region.  This is longer than the chirp ramp duration, so atoms on the leading edge and trailing edges of the pulse fall outside the chirped BCF profile and are not slowed. This effect is more pronounced with faster frequency ramps and results in a much reduced number of slowed atoms. In a fully realized design, rapidly repeating chirp sequences can be used in conjunction with a cw atomic beam to assure that most of the atoms in a continuous beam experience the full chirped BCF force, as discussed in Sec.~\ref{sec:FullSlower}.

\section{Considerations for practical atomic deceleration for MOT loading\label{sec:FullSlower}}

While our initial attempts to slow He* atoms using the chirped BCF scheme described in Sec.~\ref{sec:Chirp_Exp} have limited to $\Delta v\lesssim 370$~m/s, this is already a significant improvement, and the results have allowed us to identify straightforward improvements that will greatly extend the range.  We have also been able to verify the accuracy of the Monte Carlo model discussed in Sec.~\ref{sec:ChirpModel}.  A detailed account is presented in Chapter 7 of Ref. \cite{Chieda12}.

\begin{figure}
\includegraphics[width=\columnwidth]{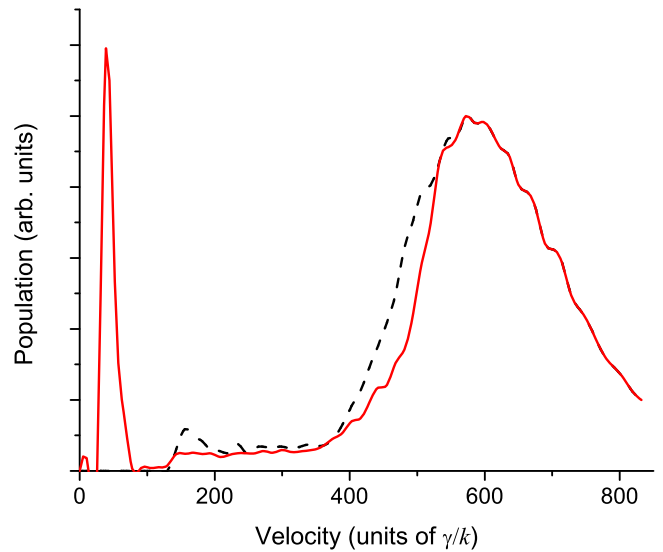}
\caption{(color online). Monte Carlo modeling for the velocity distribution of a fully realized chirped BCF slower acting on a cw atomic beam, with $\delta=123 \gamma$ and a 600 MHz chirp with a ramp duration of 10~$\mu$s. The black dashed line is the velocity distribution of the atomic beam source, and the red solid line is the predicted distribution with the chirped BCF present. Notice the large slow atom peak at $\sim 50\gamma/k$ and the relatively uniform depletion of a large range of atomic velocities.}
\label{fig:FinalModel}
\end{figure}

Even without a full redesign, simply scaling the demonstration experiment at $\delta=74\gamma$ up to a detuning of approximately $123\gamma$ and better stabilizing the rf phase will permit a doubling of the frequency chirp magnitude to 600~MHz. To accommodate the $\delta^2$ scaling of the laser power while also better matching the bichromatic beam to atomic beam diameters, about 0.5--1 W will be required from each of the two lasers.  This power level is readily available from commercially available tapered amplifier diodes, which unfortunately were not available to us at the time of the experiments.  As shown in Fig. \ref{fig:FinalModel}, Monte Carlo modeling indicates that the 600~MHz chirp, together with the increased static velocity range at a detuning of $123\gamma$, is sufficient to slow atoms from the lower half of the initial velocity distribution to final velocities of 0-100 m/s.

 With our present 160 Hz beam chopping rate, the average brightness of the slowed atomic beam would be very low due to the atom beam duty cycle of only 1\%. To create a slower capable of replacing a Zeeman slower, the chirp configuration must be adapted to use a continuous or near-continuous atomic beam. This requires that the frequency chirping cycle repeats continuously and as quickly as possible. Using a chirp ramp duration of 10~$\mu$s with an additional 5~$\mu$s to reset the laser frequency gives a cycle period of 15~$\mu$s. However, because the 5~$\mu$s is much less than the atomic transit time through the slower, it will not appreciably interfere with the cw operation of the slower. The laser frequency locking must occasionally be re-established, but it suffices to do this once every few milliseconds. For these estimates, we will assume the laser locks require 500~$\mu$s to reset, which must occur every 2~ms. This results in a fairly realistic estimate of an 80\% duty cycle.

We estimate the output flux of such a slower to facilitate comparisons to other atomic beam slowing methods. Assuming conservatively that 10\% of the metastable helium atoms are subject to slowing after taking into account the beam overlaps, BCF velocity range, and initial Doppler shift, we can predict the intensity of slow atoms using information from the model simulation shown in Fig.~\ref{fig:FinalModel}. From Sec.~\ref{sec:LDS_Exp} the metastable source brightness is $3\times10^{13}$~He*~s$^{-1}$~Sr$^{-1}$. Adjusting for the duty cycle and slower efficiency, the BCF slowed atom brightness will therefore be $0.8\times0.1(3\times 10^{13}~\text{He*~s}^{-1}~\text{Sr}^{-1})=2.4\times10^{12}$~He*~s$^{-1}$~Sr$^{-1}$. Factoring in the measured acceptance angle of the slower and its estimated exiting beam diameter of 500~$\mu$m, the estimated slow atom flux is $(2.4\times10^{12}$~He*~s$^{-1}$~Sr$^{-1})(1.5\times10^{-5}~\text{Sr})/0.2~\text{mm}^2 \simeq 2\times10^8~\text{He* s}^{-1}\text{mm}^{-2}$.

This can be compared with the 1083 nm Zeeman slower used by Vassen group, which has a MOT loading time of 0.5~s~\cite{Molenaar97, Koelemeij03, Tol99, Tychkov04, Rooijakkers96, Rooijakkers97, Tol05}. Their scheme begins with a dc discharge He* source somewhat similar to ours, but with a higher brightness of $3\times10^{14}$~He*~s$^{-1}$~Sr$^{-1}$.  A two-part Zeeman slower \cite{Tol99,Rooijakkers97} is used in conjunction with an optical molasses pre-collimator that increases the angular acceptance of the slower by a factor of $\sim30$~\cite{Rooijakkers96}. The output flux of slow He* atoms is $2\times10^9$~He*~s$^{-1}$~mm$^{-2}$~\cite{Tol05}.  Taking into account the brighter initial metastable helium source, this is the same flux that would be predicted for a chirped BCF decelerator without a pre-collimator.  The main reasons for the improved performance of the BCF slower are the greatly reduced length and the absence of transverse heating effects common in long Zeeman slowers.  With the addition of a pre-collimator the output flux could be further increased by a factor of 2--50, depending on the method used~\cite{Molenaar97, Joffe93, Partlow04}.

Development of an improved chirped slower able to slow a cw atomic beam to rest is currently underway in our laboratory.

\section{Summary\label{sec:Summary}}
Our investigation of the bichromatic force at large detunings shows that a static two-stage slower for a metastable helium beam is probably not feasible.  We have added to prior experimental evidence for an effective upper limit to the BCF at a detuning of about $250\gamma$ for He*, probably due to accumulated small dephasings of the stimulated Rabi cycling in between ``resets" by radiative decay. In addition, it is difficult to reliably produce the required laser irradiance at larger detunings because it increases quadratically.

To circumvent these limits a novel BCF slower was developed, using a small detuning while chirping the Doppler shift offset frequencies to maintain resonance with the decelerating atoms. In a prototype experiment we have achieved slowing of metastable helium by $210~\gamma/k$ (370 m/s) with a detuning of only $74\gamma$. This is more than twice the largest BCF slowing previously attained, and represents a savings in laser irradiance by nearly a factor of 10. Scaling up the detuning to $123\gamma$, increasing the bichromatic beam waists, and rapidly repeating the chirp will allow the realization of a He* slower capable of slowing atoms to MOT capture velocities. The chirped BCF slower is predicted to have a brightness at least comparable to current Zeeman slowers, but with a much shorter length.

\begin{acknowledgments}
Financial support was provided by the University of Connecticut Research Foundation and the National Science Foundation.  We thank Scott Galica for valuable contributions to numerical modeling of the BCF in the presence of unbalanced laser irradiances.
\end{acknowledgments}

\end{document}